\documentstyle[12pt]{article}
\begin{document}
\def\beq{\begin{equation}}
\def\eeq{\end{equation}}
\def\bea{\begin{eqnarray}}
\def\eea{\end{eqnarray}}
\def\ve{\vert}
\def\nnb{\nonumber}
\def\ga{\left(}
\def\dr{\right)}
\def\aga{\left\{}
\def\adr{\right\}}
\def\rar{\rightarrow}
\def\nnb{\nonumber}
\def\la{\langle}
\def\ra{\rangle}
\def\ba{\begin{array}}
\def\ea{\end{array}}

\title{ {\small {   \bf   EXCITED HEAVY MESONS DECAY 
                          FORMFACTORS IN LIGHT CONE  QCD   } } }

\author{ {\small T. M. ALIEV \thanks {e-mail:taliev@rorqual.cc.metu.edu.tr}\,\,, 
A. \"{O}ZP\.{I}NEC\.{I} \thanks
{e-mail:e100690@orca.cc.metu.edu.tr} \,\, and \,\,
M. SAVCI \thanks {e-mail:savci@rorqual.cc.metu.edu.tr}} \\ 
{\small Physics Department, Middle East Technical University} \\ 
{\small 06531 Ankara, Turkey} }

\begin{titlepage}
\maketitle
\thispagestyle{empty}

\begin{abstract}
\baselineskip  0.7cm
The formfactors of $B_1 \rar \pi$ and $D_1 \rar \pi$ transitions,
where $B_1(D_1)$ is the $1^+$ P-wave $\bar Q \gamma_{\nu}\gamma_5q$
meson state, are calculated in the framework of the 
light cone QCD. Furthermore these formfactors are compared with the
pole dominance model prediction using the values of the strong
coupling constants $g_{B_1B^*\pi}$ and $g_{D_1D^*\pi}$, and a
good agreement between these two different descriptions is observed.  

\end{abstract}

\vspace{1cm}
~~~PACS numbers: 13.20.He, 13.25.Hw, 13.30.Eg

\end{titlepage}
\baselineskip  .7cm
\newpage

\setcounter{page}{1}
\section{Introduction}
Understanding  the formfactors of the hadronic 
currents is one of the main problems in particle physics. As is well 
known, the knowledge of the formfactors allows one the determination 
of the quark mixing parameters, in a unique way \cite{R1}. Moreover, 
it gives us a direct information about the dynamics of the 
hadronic processes at large distance, where perturbation theory of
QCD is invalid. At present there exists no theoretical method
for calculating the formfactors, starting from a fundamental QCD
Lagrangian. Therefore, in estimating the formfactors, usually
phenomenological or semiphenomenological methods are used.

Along these lines, there exists two well known methods for  
calculating the hadronic form factors. First of these is the 
pole model description based on the vector dominance 
idea, that suggests  a momentum dependence dominated by the nearest 
pole, and the other one is the QCD sum rules method. The 
pole model approach does not have its roots on the basic 
principles of the theory and is purely phenomenological. Generally, 
it is assumed that the vector dominance approximation is valid 
at zero recoil, that is at $p^2 \rar m^2$, where $p^2$ is the square
of the momentum transfer, that depends on the formfactor, and $m$ is
the mass of the nearest vector meson. 
However, there are no reliable arguments in favour of the   
pole model, for it to be also valid at small $p^2$, that are
interesting for practical applications. 
The effective heavy quark theory predicts somewhat larger region of 
validity, characterized by $(m^2-p^2)/m_Q \sim (1~GeV)$, where $m_Q$ 
is the heavy quark mass (for more detail see \cite{R2,R3} 
and the references cited therein).
It is shown in  \cite{R4, R5} that 
the pole behaviour of the formfactors is consistent
with the $p^2$ dependence at $p^2 \rar 0$ predicted by the sum 
rule. This result has been confirmed independently by the calculations
carried within the framework of the light cone sum rules
\cite{R2, R6, R7}.  

In contrary to the
pole model, the QCD sum rules  method is based on the first principles
and on the fundamental QCD Lagrangian. In this work we employ 
an alternative version of the QCD sum rules, namely 
light cone sum rules method for the calculation of the
formfactors. We note that using this approach, the formfactors of the
$\pi A^0 \gamma^*$ transition \cite{R8}, the pion formfactors
at the intermediate momentum transfer \cite{R9}, semileptonic
B and D decays \cite{R10}, radiative $B \rar K^* \gamma$
decay formfactor \cite{R11} and $B^* \rar B \pi,~ D^* \rar D \pi$
transition formfactors \cite{R3}
are calculated. The results of all
works that have been devoted to the calculation of the formfactors confirm                             
that, the predictions of the pole model are compatible with that 
of the QCD sum rules method. Note that, in all these works the 
transition formfactors between the ground state mesons are considered.   

In connection with these observations, there follows an immediate question
to whether the same agreement holds for the prediction of the 
formfactors resulting
from the decays of the excited meson states. This article is devoted to find
an answer to the question 
cited above, more precisely, we calculate the formfactors for 
the excited $1^+~P$ wave decays $P_1 \rar P^* \pi$, where 
$P_1(B_1, D_1)$ is the excited $1^+~P$ wave meson state, and $P^*$ is the
$1^-$ state.
This article is organized as follows: In sect.2, we derive the sum rules
for the $P_1 \rar P^* \pi$ transition formfactors. Sect.3 is 
devoted to the numerical analysis and discussions.

\section{Sum Rules for the $P_1 \rar P^* \pi$ transition formfactors}
For calculation of the sum rules formfactors
for the $P_1 \rar P^* \pi$ transition, we
consider the following correlator,
\beq
\Pi_{\mu\nu}(q_2,q) =i \int d^4x~e^{iq_2x}~\la \pi^-(q) \ve
\bar d \ga x \dr \gamma_{\mu} Q \ga x \dr 
\bar Q \ga 0 \dr \gamma_{\nu} \gamma_5 u \ga 0 \dr
\ve 0 \ra~,\label{1}
\eeq
where, $\bar d \gamma_{\mu} Q$ and $\bar Q \gamma_{\nu} \gamma_5 u$ are 
the interpolating currents of $1^-$ and $1^+$, respectively, $Q$ is
the heavy quark, $q_2$  is the momentum of the $1^-$ meson
and $q$ is the pion's momentum.
When the pion is on the mass shell, i.e., $q^2=m^2_{\pi}$, the correlator
depends on two invariant variables $q_2^2$ and $(q_2+q)^2$. In further
calculations, we will set $m_{\pi}=0$.   

We start by considering the physical part of eq.(\ref{1}).
Before  calculating the physical part of the eq.(1), few words are in order. 
According to the value of angular momentum of the light degrees of
freedom $(s_l^P = {\frac{1}{2}}^+,{\frac{3}{2}}^+)$, the heavy 
quark effective theory \cite{R12, R13}, predicts the 
existence of two multiplets, each including  the $1^+$ meson:
the first one contains $(0^+, 1^+)$ and the second $(2^+, 1^+)$
mesons. In terms of the conventional $ ^{(2 s+1)}P_J$ states,
the $1^+$ states defined above (physical states), are given by the
following linear combinations \cite{R12, R14}:
\bea
\ve 1^+_{\frac{3}{2}} \ra &=& \sqrt{\frac{2}{3}} \ve ^1P_1 \ra 
                             +\sqrt{\frac{1}{3}} \ve ^3P_1 \ra   \nnb \\
\ve 1^+_{\frac{1}{2}} \ra &=& \sqrt{\frac{1}{3}} \ve ^1P_1 \ra
                             - \sqrt{\frac{2}{3}} \ve ^3P_1 \ra ~ . \nnb
\eea
Therefore, the physical part of the eq.(1) must contain the 
contributions from both of the $1^+$ mesons, and it can be written
as 
\bea
\Pi_{\mu\nu}= \sum_{l=\frac{1}{2},\frac{3}{2}}
 \frac {\la  \pi^-(q) \ve \bar d \gamma_{\mu} Q \ve 1_l^+ \ra
 \la 1_l^+ \ve \bar Q \gamma_{\nu} \gamma_5 u \ve 0 \ra }
{(q_2+q)^2-m_{1_l}^2}~. \label{2}
\eea

In \cite{R15, R16}, it was shown that in the heavy quark limit,
$m_Q>\!\!>m_q$,
\bea
\la1^+_{\frac{3}{2}} \ve \bar Q \gamma_\nu \gamma_5 u \ve 0 \ra =0 \nnb~.
\eea
The full theory can only bring small corrections to this result.
For this reason the contribution of the $1^+_{\frac{3}{2}}$ meson, in
eq.(2), can safely be neglected. Therefore, in future analysis we will
consider the contribution of the $1^+_{\frac{1}{2}}$ meson only,
and for simplicity we denote it as $P_1$.

The matrix elements entering in eq.(\ref{2}) are defined in the standard manner:
\bea
\la P_1 \ve \bar Q \gamma_{\nu} \gamma_5 u \ve 0 \ra &=&
- i f_{P_1} m_{P_1} \epsilon_{\nu}(p)~, \nnb \\  
\la \pi \ve \bar  d \gamma_{\nu} Q \ve P_1 \ra &=&F_0 \epsilon_{\nu}(p) 
+ F_1(q \epsilon) (p+q)_{\nu}+ F_2(q \epsilon) (p-q)_{\nu}~, \label{3}
\eea
where $f_{P_1}$, $\epsilon_{\nu}$ and $m_{P_1}$ are the leptonic decay 
constant, vector polarization and mass of the excited $1^+$ meson, 
respectively. $F_0$, $F_1$ and $F_2$ are the formfactors that
describe the $P_1 \rar \pi$ transition. The contributions of the 
coefficients of $F_1$ and $F_2$ to the decay rate are proportional to
\bea
 m_{\pi}^2 + \ga m_{P_1}-m_{P^*} \dr^2   ~. \nnb
\eea
From this expression it is obvious that, their contributions
are small in comparison to the $F_0$'s contribution. Our numerical
analysis shows that the contributions of $F_1$ and $F_2$ to the decay width
constitute about $20\%$ of $F_0$, and hence,
we shall neglect such terms.
Thus,    
\bea
\Pi_{\mu\nu} = - i \, \frac { f_{P_1} m_{P_1} }{(q_2+q)^2-m^2_{P_1}}
F_0 \ga -g_{\mu \nu} + \frac {p_\mu p_\nu}{m_{P_1}^2} \dr~, \nnb
\eea
and in  further analysis we will pay our attention to the structure 
$g_{\mu \nu}$ and denote its coefficient by $\Pi$. 
The physical part of the above relation takes the 
following form:
\beq
\Pi\ga q_2^2,(q_2+q)^2 \dr = -i \frac{f_{P_1} m_{P_1} F_0(q_2^2)}
{m^2_{P_1}-(q_2+q)^2} + i \int_{s_0}^{\infty}
\frac{\rho^h(q_2^2,s) ds}{s-(q_2+q)^2}~. \label{4}
\eeq
The first term on the right hand side is
the pole term due to the ground state in the heavy channel, 
while the continuum and the higher order states are taken into 
account by the dispersion integral above the threshold $s_0$.

Let us now consider the theoretical part of the correlator (\ref{1}).
This correlator function can be calculated in QCD at  
large Euclidean momenta $(q_2+q)^2$. In this case the virtuality of the 
heavy quark is of the order $m_Q^2-(q_2+q)^2$, and this is a large 
quantity. Therefore, we can expand the heavy quark propagator in powers of 
the slowly varying fields residing in the pion, that acts as the external 
field on the propagating heavy quark. The expansion in powers of
the external fields is also the expansion of the propagator in 
powers of a deviation from the light cone at $x^2=0$. The leading
contribution is obtained by using the free heavy quark propagator in
eq.(\ref{1})
\bea
S_Q^0(x) = i \int \frac{d^4x}{(2\pi)^4} \, e^{-ikx} \frac{(\not\!k+m_Q)}
{k^2-m_Q^2}~. \nnb
\eea
Hence,
\beq
\Pi_{\mu\nu}(q_2,q) = \int \frac{d^4x d^4k~}{\ga 2 \pi \dr^4}
\frac{e^{i\ga q_2-k \dr x}}
{\ga m_Q^2-k^2 \dr} \, \la\pi(q) \ve \bar q(x) \gamma_{\mu}
\ga \not\!k+m_Q \dr \gamma_{\nu} \gamma_5 u(0)\ve 0 \ra~. \label{5}
\eeq
We point out that, in eq.(\ref{5}) and throughout the course of the
present analysis, the path-ordered factor 
$P~exp\ga ig_s \int_{0}^{1}x_\mu A^{\mu}\ga ux \dr du \dr$ 
has been omitted, since in the Fock-Schwinger gauge 
$x_\mu A^{\mu}=0$, this factor is trivial.
It follows from eq.(\ref{5}) that, the answer is expressed via the pion
matrix element of the gauge invariant, nonlocal operator with a light cone 
separation $x^2 \simeq 0$. Following \cite{R17}, the two particle pion wave 
functions are defined as:
\bea
\la \pi(q) \ve \bar d \ga x \dr \gamma_{\mu}\gamma_5 u(0) \ve 0 \ra &=&
-i q_{\mu} f_{\pi} \int_{0}^{1} du~e^{iqux}\Big{[} \varphi_{\pi}(u) +
x^2 g_1(u) \Big{]}+ \nnb \\
&&+ q_{\mu} f_{\pi} \ga x_{\mu} - \frac{x^2 q_{\mu}}{q x}
\dr \int_{0}^{1} du~e^{iqux} g_2(u)~, \label{6}\\
\la \pi(q) \ve \bar d \ga x \dr \gamma_5 u(0) \ve 0 \ra &=&
\frac{f_{\pi} m_{\pi}^2}{m_u+m_d} \int_{0}^{1} du~e^{iqux} \varphi_{P}(u)~,
\label{7}
\eea
where $\varphi_{\pi}(u)$ is the leading twist $\tau = 2$, $\varphi_{P}(u)$
is $\tau = 3$ and $g_1(u),~g_2(u)$ are the $\tau = 4$ two particle
pion wave function. Using eq.(\ref{6}) and eq.(\ref{7}) after 
integrating over $x$ and $k$, we get for the structure $g_{\mu \nu}$:
\bea
\Pi\ga q_2,q \dr&=& -i~\Bigg\{ \frac{m_{\pi}^2}{m_u+m_d} m_Q
\int_{0}^{1} du \frac{\varphi_P \ga u \dr}{\Delta}
+2 m_Q^2 \int_{0}^{1} du \frac{g_2 \ga u \dr}{\Delta^2}-  \nnb \\
&&
- q_2 q \Bigg{[} 4 \int_{o}^{1} du \frac{\ga g_1 \ga u \dr + G_2 \ga u \dr
\dr}
{\Delta^2} 
\ga 1 + \frac{2 m_Q^2}{\Delta} \dr -
\int_{0}^{1} du \frac{\varphi_{\pi} \ga u \dr}{\Delta}
\Bigg{]} \Bigg\}~, \label{8}
\eea
where,
\bea
\Delta &=& m_Q^2-\ga q_2+qu \dr^2~, \nnb \\
G_2(u) &=& -\int_{0}^{u} g_2(v) dv~. \nnb
\eea

Since the higher order twist contributions are taken into account
in eq.(\ref{8}), the terms up to $\tau=4$, which comes from the propagator 
expansion, must also be considered.
The complete expansion of the propagator is given
in \cite{R18}, and it contains the contributions from the nonlocal
operators $\bar q G q,~\bar q GG q~$, and $\bar q q \bar q q$.
The only operator we consider here is $\bar q G q$, because
the contributions from the other two are very small, and hence they
are neglected (for more detail see \cite{R17, R18}).
Under these approximations, the heavy quark propagator
is given by the following expressions \cite{R3, R17}:
\bea
S_Q(x)& =& S^0_Q(x) - i g_s \int \frac{d^4k}{(2 \pi)^4} \, e^{-ikx}
\int_{0}^{1}du \Bigg{[}~ \frac{1}{2}~ \frac{\not\!k+m_Q}{\ga m_Q^2-k^2
\dr^2}~
G^{\mu \nu}(ux) \sigma_{\mu \nu}+ \nnb \\
&&+ \frac{1}{m_Q^2-k^2}~ u x_\mu G^{\mu
\nu}(ux) \gamma_\nu \Bigg{]}~. \label{9}
\eea  
Substituting second term in eq.(\ref{9}) into eq.(\ref{1}), we get,
\bea
\Pi_{\mu\nu}(q_2,q) &=& \int \frac{d^4x~d^4k~du}{(2 \pi)^4}
\frac{e^{i(q_2-k)x}}{\ga m_Q^2-k^2 \dr^2}~ \la \pi(q) \ve 
\bar d(x) \gamma_\mu \Big{[}
u x_\rho G^{\rho \lambda}(ux) \gamma_\lambda + \nnb \\
&& +\frac{\not\!k+m_Q}{\ga m_Q^2-k^2 \dr^2}~ \frac{1}{2}~ G^{\rho \lambda}(ux)
\sigma_{\rho \lambda} \Big{]} \gamma_\nu \gamma_5 u(0) \ve 0 \ra~ . \label{10}
\eea
Using the following identities,
\bea
\gamma_\mu \gamma_\lambda \gamma_\nu &=& g_{\mu\lambda} \gamma_\nu
+g_{\lambda\nu} \gamma_\mu - g_{\mu\nu} \gamma_\lambda
-i \epsilon_{\mu\lambda\nu\tau} \gamma^\tau \gamma_5~, \nnb \\
\gamma_\mu \sigma_{\rho \lambda} &=& i \ga g_{\mu\rho} \gamma_\lambda    
-g_{\mu\lambda} \gamma_\rho \dr + \epsilon_{\mu\rho\lambda\tau} \gamma^\tau
\gamma_5~, \nnb
\eea
and the definition of the three particle pion wave functions \cite{R17},
\bea
&&
\la \pi(q) \ve \bar d(x) \gamma_{\mu}\gamma_5 g_s G_{\alpha \beta}(ux) u(0)
\ve 0 \ra = \nnb \\
&& f_{\pi} \Bigg\{ \Bigg{[} q_{\beta} \ga g_{\alpha \mu} - \frac {x_{\alpha}
q_{\mu}}
{qx} \dr - q_{\alpha} \ga g_{\beta \mu} \frac {x_{\beta} q_{\mu}}{qx} \dr
\Bigg{]} \int {\cal D}\alpha_i \varphi_{\perp}(\alpha_i) \, e^{iqx(\alpha_1+u
\alpha_3)}+ \nnb \\
&& +\frac{q_{\mu}}{qx} \ga q_{\alpha} x_{\beta} - q_{\beta}
x_{\alpha} \dr
\int {\cal D}\alpha_i \varphi_{\parallel}(\alpha_i) \, e^{iqx(\alpha_1+u
\alpha_3)} \Bigg\}~ \label{11}~,
\eea

\bea
&&
\la \pi(q) \ve \bar d(x) \gamma_{\mu}g_s \tilde{G}_{\alpha \beta}(ux) u(0)
\ve 0 \ra = \nnb \\
&& i f_{\pi}\Bigg\{  \Bigg{[} q_{\beta} \ga g_{\alpha \mu} - \frac {x_{\alpha}
q_{\mu}}
{qx} \dr - q_{\alpha} \ga g_{\beta \mu} \frac {x_{\beta} q_{\mu}}{qx} \dr
\Bigg{]} \int {\cal D}\alpha_i \tilde{\varphi}_{\perp}(\alpha_i)
\, e^{iqx(\alpha_1+u \alpha_3)}+ \nnb \\
&&+\frac{q_{\mu}}{qx} \ga q_{\alpha} x_{\beta} - q_{\beta}
x_{\alpha} \dr
\int {\cal D}\alpha_i \tilde{\varphi}_{\parallel}(\alpha_i)
\, e^{iqx(\alpha_1+u \alpha_3)} \Bigg\}~, \label{12}
\eea
we obtain from eq.(\ref{10}), for  the structure $g_{\mu\nu}$
\beq
\Pi=i q_2 q \int_{0}^{1} du \int {\cal D}\alpha_i 
\frac{(1-2u)\varphi_{\parallel}(\alpha_i) +\tilde{\varphi}_{\parallel}(\alpha_i)}
{\left[ m_Q^2 - \left[ q_2+q_1(\alpha_1+u \alpha_3) \right]^2 \right]^2}~,
\label{13}
\eeq
where, ${\cal D}\alpha_i \equiv d{\alpha_1}d{\alpha_2}d{\alpha_3}
\, \delta(1-\alpha_1-\alpha_2-\alpha_3)$, and 
$\tilde{G}_{\alpha\beta}=\frac{1}{2}\epsilon_{\alpha\beta\mu\nu}G^{\mu\nu}$,
 is the dual tensor to $G^{\mu\nu}$, and $\varphi$'s are all twist four
fuctions. Collecting all the terms (eqs.(\ref{8}) and (\ref{9}))for 
the theoretical part of the of the correlator function (\ref{1}), we get,
\bea
\Pi^{\it theor}&=&
-i \Bigg\{  
\frac{m_{\pi}^2}{m_u+m_d} m_Q \int_{0}^{1} du \frac{\varphi_P(u)}{\Delta}+
2 m^2_Q \int_{0}^{1} du \frac{g_2(u)}{\Delta^2}+ \nnb \\
&&+ q_2 q \Bigg{[} \int_{0}^{1} du \frac{\varphi_{\pi}(u)}{\Delta}
- 4 \int_{0}^{1} du \frac{(g_1(u)+G_2(u))}{\Delta^2} 
\ga 1+ \frac{2 m_Q^2}{\Delta} \dr \Bigg{]} - \nnb \\
&&- q_2 q \int_{0}^{1} du \int {\cal D}\alpha_i
\frac{(1-2u)\varphi_{\parallel}(\alpha_i)
+\tilde{\varphi}_{\parallel}(\alpha_i)}
{\Delta_1^2} \Bigg\}~, \label{14}
\eea
where,
\bea
\Delta_1 &=& m_Q^2-\left[ q_2+ q \ga \alpha_1+u \alpha_3 \dr \right]^2~. \nnb
\eea
The sum rule for the formfactor $F_0(q_2^2)$ is obtained by equating
the expressions (\ref{4}) and (\ref{14}) for the invariant amplitude 
$\Pi\ga q_2^2,(q_2+q)^2 \dr$, which results as:
\bea
\frac{f_{P_1} m_{P_1} F_0(q_2^2)}
{m^2_{P_1}-(q_2+q)^2} + \int_{s_0}^{\infty}
\frac{\rho^h(q_2^2,s) ds}{s-(q_2+q)^2} = \Pi^{theor}. \nnb
\eea
Invoking the duality prescripton and performing the  Borel
transformation in the variable $(q_2+q)^2$, we get the following 
sum rules for the form factor $F_0(q_2^2)$ :
\bea
F_0(q_2^2) &=& \frac{f_\pi}{f_{P_1}m_{P_1}}
\Bigg\{
\int_{\delta}^{1} \frac{du}{u}~ exp \left[ \frac{m_{P_1}^2}{M^2}
-\frac{m_Q^2-q_2^2 \bar u}{u M^2} \right]~ \Phi_2 (u,M^2,q^2) + \nnb \\
&& + \frac{1}{2} \int_{0}^{1} du \int \frac{{\cal D}\alpha_i}
{\ga \alpha_1+u \alpha_3 \dr^2}\Theta (\alpha_1 + u \alpha_3 - \delta)
\times \nnb \\
&&\times exp \left[ \frac{m_{P_1}^2}{M^2}-\frac{m_Q^2-q_2^2 \bar u}{u M^2}\right]~
\Phi_3 (u,M^2,q^2) 
\Bigg\}~, \label{15}
\eea
where,
\bea
\delta &=& \frac{m_Q^2-q_2^2}{s_0-q_2^2}~, \nnb\\
{\bar u} &=& 1-u~, \nnb
\eea
and,
\bea
\Phi_2 &=& \frac{m_\pi^2}{m_u+m_d} m_Q \varphi_P(u) +
\frac{m_Q^2-q_2^2}{u} \varphi_{\pi}(u) + \nnb \\
&& + 2\frac{g_1(u)}{u} 
\left[1 + \frac{m_Q^2-q_2^2}{u M^2} - 
\frac{m_Q^2(m_Q^2-q_2^2)}{u^2 M^4}  
\right] + \nnb \\
&& + 2\frac{G_2(u)}{u}\left[ 1-\frac{m_Q^2-q_2^2}{u M^2} \right]~, \nnb \\
\Phi_3 &=& \left[ (1-2u) \varphi_{\parallel}(\alpha_i)+
{\tilde{\varphi}_{\parallel}(\alpha_i)} \right] 
\ga 1- \frac{m_Q^2-q_2^2}{\ga \alpha_1 + u \alpha_3 \dr M^2 } \dr~. \nnb
\eea 
\section{Numerical analysis and discussion}
Now we turn to the numerical calculations. In expression (\ref{15}),
we use the following set of parameters: $m_{B_1}=5.738~GeV,~
m_{D_1}=2.4~GeV,~(s_0)_B= 35~GeV^2,~(s_0)_D= 8~GeV^2$. For leptonic
decay constants $f_{B_1}$ and $f_{D_1}$, we use the results given in
\cite{R19}: 
$f_{B_1}\simeq 0.2\pm 0.02~GeV$ and $f_{D_1}\simeq 0.3\pm 0.03~GeV$. 
The highest value of the Borel Parameter $M^2$ is fixed by imposing
the condition that the continuum contribution is $30\%$ of the 
resonance. With the help of this restriction, we calculate the maximum
value of the Borel parameter for $B(D)$  to be
$M^2_{max} \simeq 20~GeV^2(\simeq 8~GeV^2)$. The minimal value of
the same parameter is usually fixed by the condition that the 
terms proportional to the higher powers of $\frac{1}{M^2}$ are negligible 
and it is found to have the value 
$M_{min}^2 \simeq 8~GeV^2(\simeq 2~GeV^2)$ for $B(D)$ mesons.
Our calculations show that, the variation of $M^2$ within the 
above-mentioned limits, changes the result by less than $10\%$, which
means that the dependence of the formfactor $F_0(q_2^2)$ on the 
Borel paremeter $M^2$ is quite weak. Note that, a similiar 
situation exists for the $B \rar \pi$ and $D \rar \pi$ transitions
too (see for example \cite{R2, R3}). Because of that, in our analysis
we take $M^2 = 15~GeV^2$ for the $B_1 \rar \pi$ and 
$M^2 = 4~GeV^2$ for the $D_1 \rar \pi$ transition formfactors.

The maximum momentum transfer $q_2^2$ at which the sum rule (\ref{15})
is applicable, is  about $15-20~GeV^2$ for B meson and
$1~GeV^2$ for the D meson cases, respectively(for more detail see
\cite{R2,R3}).
The explicit form of the pion wave function, used in eq.(\ref{15})
can be found in \cite{R17}. The momentum transfer $q_2^2$ dependence
of the formfactors $F_0^B(q_2^2)$ and $F_0^D(q_2^2)$ are 
plotted in Fig.1 and Fig.2. From these figures
we observe that, 
\bea
F^B_0(q_2^2=0) = 1.2~, \nnb \\
F^D_0(q_2^2=0) = 1.5~. \label{16}    
\eea

Let us turn our attention to the pole model prediction for $F^B_0$ and
$F^D_0$ formfactors. In \cite{R3}, it is shown that the coupling
constants $g_{B^*B\pi}$ and $g_{D^*D\pi}$ fix the normalization
of the formfactors of the  $B \rar \pi$ and $D \rar \pi$ transitions,
respectively, within the context of pole model (see also,
\cite{R4, R5}). Using the pole model
description in a similar manner, the formfactor $F_0(q_2^2)$ can be
expressed via the $g_{P_1P^*\pi}$ coupling constant, that is 
calculated using the same correlator function (\ref{1}) in \cite{R19}.
Indeed, the formfactor $F_0(q_2^2)$ defined by the matrix element,
\bea
\la \pi(q) \ve \bar d \gamma_\mu Q \ve P_1(p) \ra = 
F_0(q_2^2) \epsilon_\mu(p)
+F_1(q_2^2)(\epsilon q) (p+q)_\mu +  
F_2(q_2^2)(\epsilon q)(p-q)_\mu~, \label{17}
\eea
is predicted to be (for the structure $g_{\mu\nu}$)
\beq
F_0(q_2^2)=\frac{g_{P_1 P^*\pi} f_{P^*}}{m_{P^*} \ga 1- 
\frac{q_2^2}{m^2_{P^*}} \dr }~. \label{18}
\eeq
In deriving eq.(\ref{18}),
we used the following definition,
\beq
\la \pi P^* \ve P_1 \ra \equiv g_{P_1 P^*\pi} 
\ga \epsilon \epsilon^* \dr~, \label{19}
\eeq
where $\epsilon$ and $\epsilon^*$ are the 4-polarization 
vectors of the $P_1$  and $P^*$ mesons, respectively.
The dependence of the formfactors  $F_0^B(q_2^2)$ and  $F_0^D(q_2^2)$ 
on $q_2^2$ in the pole model (eq.(\ref{18})) are plotted in 
Fig.1 and Fig.2, where we use $g_{B_1 B{*}\pi} = 24 \pm 3~GeV$ and 
$g_{D_1 D^{*}\pi} = 10 \pm 2~GeV$ \cite{R19}. 
From these figures we conclude that, in the overlap region
both calculations agree, approximately, with each other.
Quantitatively, at $q_2^2=0$, it follows from eq.(\ref{18}) that,
\bea
F^B_0(q_2^2=0) = 0.72~, \nnb \\
F^D_0(q_2^2=0) = 1.20~. \label{20}    
\eea
If we compare eqs.(\ref{16}) and (\ref{20}), we see that, in the region
for which $m_Q^2-q_2^2 > (1~ GeV^2)$, ($Q=b,~c$), the numerical agreement
between the two approaches is better than $20\%$ for D, but 
only within $35\%$ for the B meson case.

Finally we would like to point out that, these two approaches 
lead to absolutely different asymptotic behaviours of the relevant
formfactors. Using the  HQFT results, we get
\bea
f_{P_1} \sqrt{m} = {\hat f}_{P_1} ~,
~~~~~ f_{P^*} \sqrt{m} = {\hat f}_{P^*} ~ \nnb
\eea
and,
\bea 
g_{P_1P^*\pi} \simeq \frac{2 m}{\hat g} 
~~~~\mbox{(see also \cite{R3})}~.   \nnb
\eea
From eq.(\ref{18}) it follows that, at $m_b \rar \infty$ we get,
\beq
F_0^{B(D)} \sim ({m_{B(D)}})^{-1/2}~. \label{21}
\eeq
Then from eq.(\ref{15}) we see that, in this limit
\beq
F_0^{B(D)} \sim ({m_{B(D)}})^{-3/2}~. ~~~~\mbox{(see also \cite{R20})}  \label{22}
\eeq

In conclusion, we calculated the formfactors of the excited state 
$1^+$ mesons, namely, $B_1 \rar \pi$ and $D_1 \rar \pi$ transitions,
in the framework of the light cone QCD sum rules and  compared
our results with the pole dominance model predictions. The 
comparision elaborates that, the  agreement between the two 
descriptions is rather good. This  justification demonstrates that, 
the pole dominance model works quite good for the p-wave meson 
transitions, as it does for the s-wave ones.

\vskip 15cm

\begin{figure}
\vspace{27.0cm}
    \includegraphics{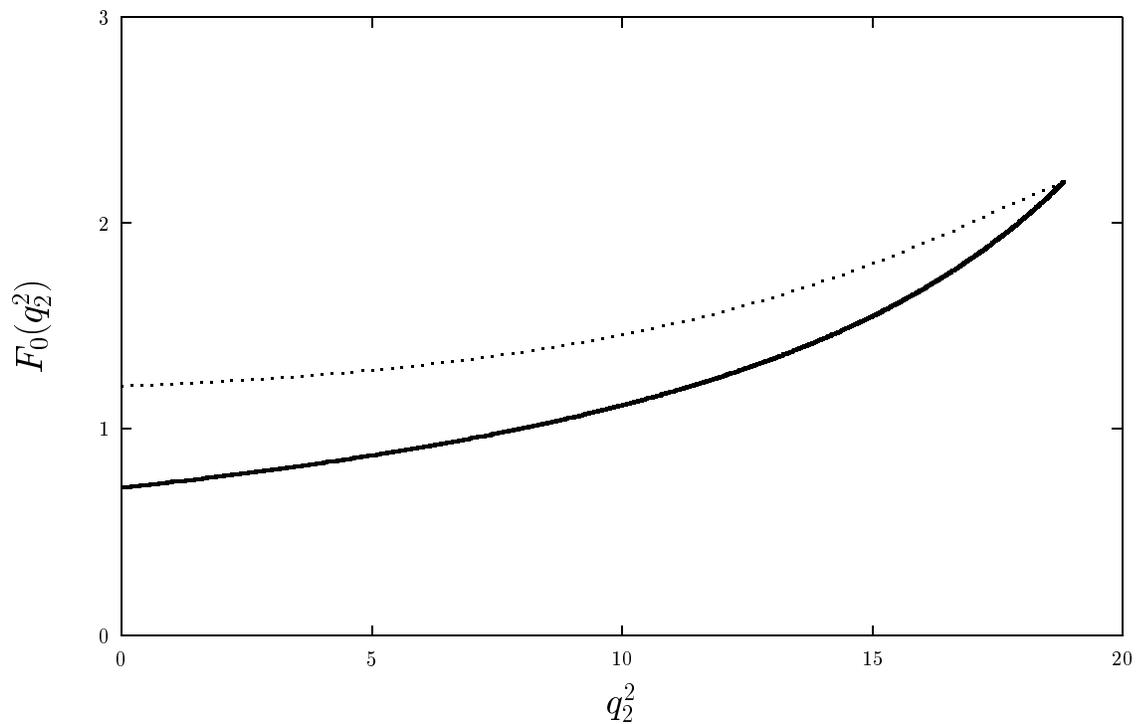}
    \vspace{-15.0cm}
                   \caption{The dependence of the formfactor $F_0(q_2^2)$,
                            for the $B_1 \rar \pi$ transition, on $q_2^2$.
                            The dotted line corresponds to the light cone QCD 
                            sum rules prediction and the solid line 
                            describes the pole dominance model prediction.  
                                                                           }
\vskip 10cm
\end{figure}

\begin{figure}
\vspace{27.0cm}
    \includegraphics{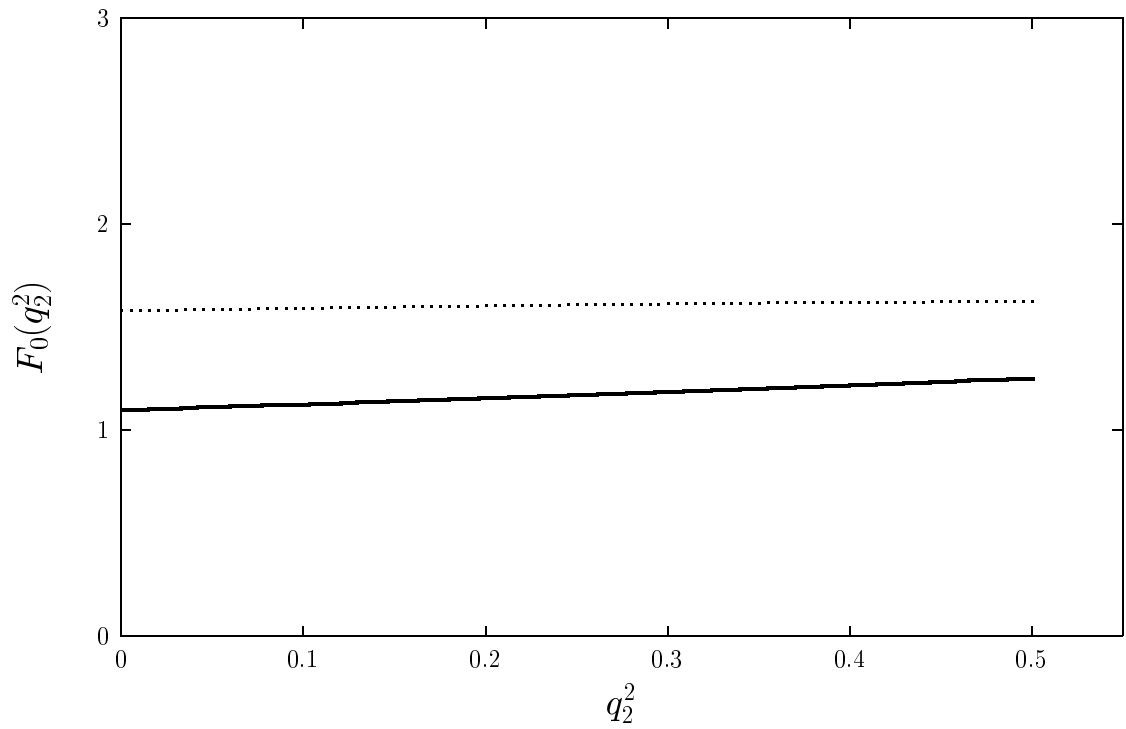}
    \vspace{-15.0cm}
                   \caption{Same as in Fig.(1), but for the  $D_1 \rar \pi$
                            transition.
                                        }
\vskip 10cm
\end{figure}


\newpage

\section*{Figure Captions}
{\bf 1.} The dependence of the formfactor $F_0(q_2^2)$,
for the $B_1 \rar \pi$ transition, on $q_2^2$.
The dotted line corresponds to the light cone QCD
sum rules prediction and the solid line 
describes the pole dominance model prediction. \\ \\
{\bf 2.} Same as in Fig.(1), but for the  $D_1 \rar \pi$
transition.



\newpage
  
\begin{thebibliography}{99}

\bibitem{R1} For a review see M. Neubert, 
{\it Phys. Rep.} {\bf C245} (1994) 259; \\
A. Ali {\it Prep. DESY} {\bf 96-106} (1996) (to be published)  

\bibitem{R2} V. M. Belyaev, A. Khodjamirian and R. R\"{u}ckl,
{\it Z. Phys.} {\bf C60} (1993) 349.

\bibitem{R3} V. M. Belyaev, V. M. Braun, A. Khodjamirian and R. R\"{u}ckl,\\
{\it Phys. Rev.} {\bf D51} (1995) 6177.

\bibitem{R4} G. Burdman, Z. Ligeti, M. Neubert and Y. Nir,
{\it Phys. Rev.} {\bf D49} (1994) 2331.

\bibitem{R5} B. Grinstein and P. F. Mende,
{\it Nucl. Phys.} {\bf B425} (1994) 451.

\bibitem{R6} P. Ball, V. M. Braun and H. G. Dosh,
{\it Phys .Lett.} {\bf B273} (1991) 316.

\bibitem{R7} P. Ball,
{\it Phys. Rev.} {\bf D48} (1993) 3190.

\bibitem{R8} V. M. Belyaev,
{\it Z. Phys.} {\bf C65} (1994) 93.

\bibitem{R9} V. M. Braun and I. Halperin, 
{\it Phys. Lett.} {\bf B328} (1994) 457.

\bibitem{R10}  P. Ball, V. M. Braun and H. G. Dosh,
{\it Phys. Rev.} {\bf D44} (1991) 3567.

\bibitem{R11} A. Ali, V. M. Braun and H. Simma, 
{\it Z. Phys.} {\bf C63} (1994) 437. 

\bibitem{R12} N. Isgur and M. B. Wise, {\it Phys. Rev. Lett.} {\bf 66}
(1991)
1130; {\it Phys. Rev.} {\bf D43} (1991) 819.

\bibitem{R13} A. F. Falk and M. Luke, {\it Phys. Lett.} {\bf B292} (1992) 119; \\
              U. Killian, J. C. K\"{o}rner and D. Pirjol, {\it ibid} {\bf
              B288} (1992) 360.

\bibitem{R14} J. L. Rosner, {\it Comm. Nucl. Part. Phys.} {\bf 16} (1986)
109.

\bibitem{R15} P. Colangelo, G. Nardulli  and N. Paver, {\it Phys. Lett.}
              {\bf B293} (1992) 207.

\bibitem{R16} R. Casalbuoni, et. al., {\it Phys. Lett.} {\bf B299} (1993)
139.

\bibitem{R17} V. M. Braun and I. B. Filyanov,
{\it Z. Phys.} {\bf C44} (1989) 157; \\ 
{\it ibid} {\it Z. Phys.} {\bf C48} (1990) 239.

\bibitem{R18} I. I. Balitsky and V. M. Braun, {\it Nucl. Phys.} {\bf B311} (1988) 541.

\bibitem{R19} T. M. Aliev, N. K. Pak and M. Savc\i,
({\it Phys. Lett.} {\bf B}, in press).

\bibitem{R20} V. L. Chernyak and I. R. Zhitnitsky,
{\it Nucl. Phys.} {\bf B345} (1990) 137.

\end{thebibliography}
\end{document}